\documentclass[sigconf]{acmart}
\setcopyright{acmlicensed}  
\copyrightyear{2026}
\acmYear{2026}
\acmConference[GREENS '26]{10th International Workshop on Green and Sustainable Software}{April 14,2026}{Rio de Janeiro, Brazil}           
\acmBooktitle{10th International Workshop on Green and Sustainable Software} 
\acmDOI{XXXXXXX.XXXXXXX} \acmISBN{} \acmPrice{}      
\usepackage{amsmath,amsfonts,booktabs,graphicx,xcolor,algorithm,algorithmic,multirow}
\usepackage[caption=false,font=footnotesize]{subfig}

\usepackage[table]{xcolor}
\definecolor{bestbg}{RGB}{230,255,230} 
\definecolor{besttext}{RGB}{255, 230, 230}   

\title{Energy Consumption of Dataframe Libraries for End-to-End Deep Learning Pipelines -- A Comparative Analysis}

\author{Punit Kumar, Asif Imran, and Tevfik Kosar}
\affiliation{
  \institution{Department of Computer Science and Engineering}
  \city{University at Buffalo, Buffalo, New York}
  \country{USA}
}
\email{{punitkum, asifimra, tkosar}@buffalo.edu}

\copyrightyear{2026}
\acmYear{2026}
\setcopyright{cc}
\setcctype{by}
\acmConference[GREENS '26]{10th International Workshop on Green and Sustainable Software }{April 12--18, 2026}{Rio de Janeiro, Brazil}
\acmBooktitle{10th International Workshop on Green and Sustainable Software (GREENS '26), April 12--18, 2026, Rio de Janeiro, Brazil}
\acmPrice{}
\acmDOI{10.1145/3786148.3788619}
\acmISBN{979-8-4007-2381-0/2026/04}

\begin{document}

\begin{abstract}
This paper presents a comparative performance analysis of three popular Python data manipulation libraries—Pandas, Polars, and Dask—within the context of deep learning training pipelines. The existing studies in this area do not embed the libraries inside a full deep-learning training pipeline where \textit{data loading}, \textit{preprocessing}, and \textit{batch feeding} interact tightly with GPU workloads.
To bridge this gap, we integrate Pandas, Polars, and Dask into representative deep learning training and inference pipelines and conduct experiments across a wide range of various machine learning models and datasets, measuring key performance indicators such as runtime, memory usage, disk usage, and energy consumption (CPU and GPU). Our comprehensive analysis reveals that Polars consistently minimizes CPU energy consumption on larger workloads, while Pandas remains competitive for moderate sizes. Dask’s overhead can lead to higher energy usage on small to moderate datasets. All three libraries achieve similar runtimes for heavy GPU workloads (ResNet, Mask R-CNN). Polars and Pandas maintain lower CPU memory footprints than Dask, but Dask offers easier scalability if data truly exceeds available RAM. Polars shows marginal energy savings on the CPU during preprocessing.
\end{abstract}


\maketitle

\section{Introduction}
Efficient data handling is crucial in the development and deployment of deep learning models. Modern deep learning workflows often involve large-scale data preprocessing, feature engineering, and iterative model training. In Python, several libraries exist to facilitate data manipulation, each offering different trade-offs between ease of use, memory footprint, parallelism, and scalability. Pandas~\cite{gupta2024introduction} is a widely adopted library known for its intuitive API and rich functionality for structured data. However, it can become a bottleneck when datasets exceed main memory or when operations are not vectorized. Polars~\cite{lutkebohmert2021polars}, a more recent entrant, leverages Apache Arrow~\cite{ApacheArrow} under the hood and employs parallel execution by default, promising significant performance improvements over Pandas for both in-memory and out-of-core tasks. Dask~\cite{rocklin2015dask}, on the other hand, provides a framework for parallel and distributed computing by scaling Pandas operations to multiple cores or even clusters, making it suitable for larger-than-memory datasets.

Several groups benchmark DataFrame backends, but mostly in isolation or with simple analytics workloads. Mozzillo \textit{et al.} evaluate seven libraries on four real-world tasks~\cite{mozzillo2023single}; Broihier \textit{et al.} propose \emph{PandasBench}, a corpus of 102 notebooks for reproducible testing~\cite{broihier2025pandasbench}; and Bergamaschi \textit{et al.} study energy-to-solution for Polars and Pandas~\cite{nahrstedt2024empirical}. None of these works embeds the libraries \emph{inside} a full deep-learning training pipeline where \textit{data loading}, \textit{preprocessing}, and \textit{batch feeding} interact tightly with GPU workloads.

To bridge this gap, we integrate Pandas, Polars, and Dask into representative pipelines for a variety of workloads, such as \textit{Random Forest} \cite{grinsztajn2022tree}, \textit{SVR}, \textit{TabNet} \cite{arik2021tabnet}, \textit{XGBoost} \cite{grinsztajn2022tree}, \textit{ResNet-50} \cite{grinsztajn2022tree}, and \textit{Mask R-CNN} across datasets ranging from an \textit{insurance policy table~\cite{insurance_kaggle}} to \textit{Common
Objects in Context (COCO) images~\cite{lin2014coco}}. We record runtime, peak CPU/GPU memory, disk I/O, and energy consumption for each implementation. CPU events are captured with Linux \texttt{perf}~\cite{perf}; GPU power is sampled via NVIDIA \texttt{pynvml}~\cite{pynvml}. Each configuration is run five times, and we report the trimmed mean and 95\% confidence intervals to suppress outliers.

\begin{figure*}[!t]
  \centering
  \includegraphics[width=0.65\textwidth]{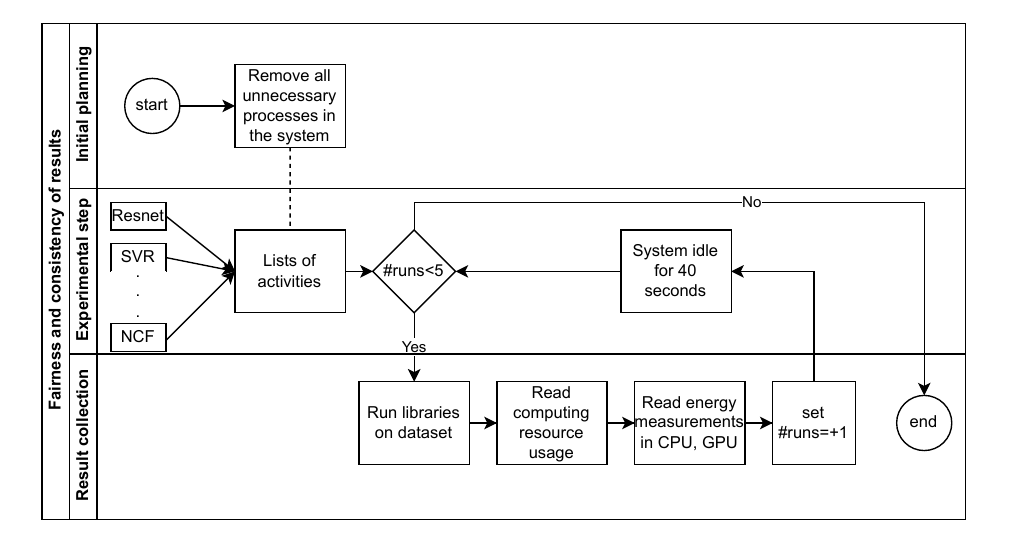}
  \caption{Flowchart to highlight the experimental process of this study.}
  \label{fig:fairness}
\end{figure*}

This paper makes novel contributions to the field by answering the following research questions:
\begin{itemize}
  \item \textbf{RQ1: How do Pandas, Polars, and Dask differ in performance and  scalability when applied to machine learning workflows?} We present a complete comparison framework of Dataloader (Pandas, Polars, and Dask) within machine learning workflows, ranging from classical models to deep neural networks. Unlike prior work that focused on isolated steps (for example, data loading or training only), we evaluate their impact across the entire training and inference pipeline.
  
  \item \textbf{RQ2: How do DataFrame libraries impact the performance of training vs inference steps of machine learning workflows?} While existing studies have examined Pandas, Polars, and Dask in isolation—often focusing solely on data loading or preprocessing—a critical gap remains: their comprehensive influence across complete ML workflows. We address this by conducting the first end-to-end comparison spanning classical models to deep neural networks, measuring performance from initial data ingestion and training through final inference.
  
  \item \textbf{RQ3: How do different dataloader setups impact CPU, GPU, RAM, and VRAM energy consumption?} We go beyond runtime analysis and include detailed measurements of CPU and GPU energy consumption. Our framework also tracks energy consumption due to memory usage (RAM and VRAM), providing a complete energy profile for each dataloader setup.
\end{itemize}

We evaluate these dataloaders on a diverse set of datasets, from small (Insurance~\cite{insurance_kaggle}, ML-1M) to large-scale COCO images and annotations (from  Microsoft COCO~\cite{lin2014coco}) and offer practical recommendations on when to use each backend based on dataset size, model type, and hardware configuration. This framework quantifies energy efficiency across data loading frameworks in the context of full ML pipelines, helping bridge the gap between performance benchmarking and sustainable machine learning.

The rest of the paper is organized as follows: Section~\ref{sec:experiments} explains the experiments, which include a comparative analysis of dataloaders and a breakdown of energy profiling. Section~\ref{sec:methodology} highlights the integration of backends into the training pipelines and describes performance metrics. Section~\ref{sec:results} presents the results of the experiments and summarizes our findings. Section~\ref{sec:relatedwork} discusses the related work, and Section~\ref{sec:conclusion} concludes the paper.

\section{Experimental Setup}
\label{sec:experiments}

\subsection{Hardware and Software Environment}
All experiments are conducted on a workstation equipped with a 16-core, 32-thread CPU clocked at 2.10 GHz, 128 GB of RAM, an NVIDIA Tesla V100 GPU with 32 GB HBM2 memory, and a 1 TB NVMe SSD, running Ubuntu 20.04 LTS with Linux kernel 5.4. The software environment is managed using Conda and includes the following key libraries and tools: \textit{Python 3.8.12, Pandas 1.4.2, Polars 0.10.18, Dask 2022.2.0, TensorFlow 2.8.0, PyTorch 1.12.1, scikit-learn 1.0.2, XGBoost 1.5.0, PyTorch-TabNet 3.1.1, the Linux perf tool (version 5.4),} and \textit{NVIDIA Management Library Python bindings (pynvml) version 11.5.50.}

Energy and power measurements during experiments can be affected by various external factors. To minimize such influence and promote consistency, each experiment is executed five times, and the average result is used, following the methodology described by Georgiou et al.~\cite{georgiou2022green}. The methodology shown in Figure \ref{fig:fairness} is designed to ensure fairness and consistency in performance and energy measurements. It begins by removing unnecessary system processes and selecting a list of activities, such as running models like ResNet, SVR, and NCF. Each activity is executed multiple times while the system records computing resource usage and energy consumption for the CPU and GPU. After every run, the system enters an idle state for 40 seconds using the "sleep" command to stabilize conditions and increments the run counter. This pause aligns with the recommendation by Bornholt et al.~\cite{bornholt2012model}, who note that allowing the system to idle for at least 30 seconds helps eliminate residual power effects and enables the system to return to a stable baseline. This process repeats until the predefined number of runs is reached, after which the experiment ends. 

\subsection{Datasets}
We select three representative datasets described below. The reason for their choice mainly depends on the fact that they are representative of real-world problems, widely recognized in the community, and well-suited to benchmark their methods \cite{xia2024contemporary, harper2015movielens}.
\begin{itemize}
    \item \textbf{Insurance Dataset~\cite{insurance_kaggle}:} A tabular dataset of 1,000 insurance policy records with features such as age, BMI, smoking status, and region. The target is primarily policy cost (regression). We include this dataset because it offers structured, real-world tabular data suitable for testing predictive models in regression and classification problems.
    \item \textbf{ML‐1m (MovieLens 1M)~\cite{harper2015movielens}:} Approximately 1,000,000 user‐item rating pairs with user and movie metadata. Used for Collaborative Filtering and NCF (Neural Collaborative Filtering) experiments. It is a standard benchmark dataset for recommender systems, providing a realistic set of user-movie interactions with metadata for reproducible evaluation.
    \item \textbf{COCO~\cite{lin2014coco}:} “Common Objects in Context” image detection dataset (COCO) consists of 80 object categories with $\sim$118,000 training images. COCO\_2017 is a subset of the COCO dataset with updated annotations which is used for ResNet (classification) and Mask R‐CNN (object detection/ segmentation) experiments.
\end{itemize}
Each dataset is stored in CSV (insurance, ML‐1m) or COCO JSON plus image directories (COCO, COCO\_2017). We generate Parquet copies where applicable to exploit Polars' Arrow‐native format, while Pandas and Dask read directly from CSV.

\section{Methodology}
\label{sec:methodology}

\begin{algorithm}[b]
\small
\caption{Unified Data Processing Pipeline}
\label{alg:data_pipeline}
\begin{algorithmic}[1]
\REQUIRE Input Data (CSV/Parquet), Framework choice ($F$): Pandas, Polars, or Dask
\ENSURE Batched Dataset ready for Training
\IF{$F$ = Pandas}
  \STATE Load data into \texttt{pandas.DataFrame};
  \STATE Preprocess (missing value imputation, categorical encoding, normalization);
  \STATE Convert DataFrame to NumPy arrays or ML Dataset object;
  \STATE Slice DataFrame into mini-batches using indices or sampling;
\ELSIF{$F$ = Polars}
  \STATE Load data into \texttt{polars.DataFrame} (lazy or eager);
  \STATE Preprocess via lazy expressions (\texttt{with\_columns}, \texttt{filter});
  \STATE Collect DataFrame and convert to NumPy or ML Dataset object;
  \STATE Perform zero-copy slicing into mini-batches;
\ELSIF{$F$ = Dask}
  \STATE Load data into \texttt{dask.dataframe.DataFrame} with multiple partitions;
  \STATE Preprocess using parallelized \texttt{map\_partitions} and \texttt{compute()};
  \STATE Convert to Pandas DataFrame (if feasible) or iterate partitions;
  \STATE Batch slicing using asynchronous scheduler;
\ENDIF
\STATE Return mini-batches for model training;
\end{algorithmic}
\end{algorithm}
\subsection{Integration of Backends into Training Pipelines}
For each library (Pandas, Polars, Dask), we implement a data loading and preprocessing module that feeds the downstream model training loop. Algorithm \ref{alg:data_pipeline} describes a unified data processing pipeline that prepares raw tabular data for machine learning training across different DataFrame frameworks, namely Pandas, Polars, and Dask. It begins by loading input data into the selected framework and then applies preprocessing steps such as missing value imputation, categorical encoding, and normalization. After preprocessing, the data is converted into a machine learning–compatible format, such as NumPy arrays or a dataset object. The pipeline then slices the processed data into mini-batches suitable for training, using framework-specific optimizations: standard indexing in Pandas, lazy evaluation and zero-copy slicing in Polars, and parallel, partition-based processing with an asynchronous scheduler in Dask. Overall, the algorithm provides a consistent and portable workflow that abstracts away framework-specific details while ensuring efficient, batched data delivery for model training.

The training loop, shared across all backends, begins by assembling each mini-batch through the retrieval of the corresponding feature and label tensors. The model then performs a forward pass to generate predictions and compute the associated loss. Next, a backward pass is executed to calculate gradients and update the model parameters through optimization. Throughout this process, key performance indicators such as training loss and validation accuracy are recorded. This structured workflow ensures consistent training behavior and comparable evaluation across different backends.

\subsection{Measurement of Performance Metrics}
Accurate measurement of resource usage and energy consumption is critical, and we employ industry-standard tools for this purpose as follows:

\subsubsection{CPU Metrics}
We use the Linux \texttt{perf} tool ~\cite{perf} to record CPU-side metrics. The \texttt{perf stat} command aggregates these counters over the entire process lifetime. We parse \texttt{perf\_output.txt} to extract:
\begin{itemize}
  \item \textbf{Runtime (s):} Wall-clock time recorded by perf.
  \item \textbf{CPU Energy (J):} Sum of \texttt{energy-cores} and \texttt{energy-ram} metrics.
  \item \textbf{CPU Memory (MB):} Approximated by peak resident set size reported.
  \item \textbf{Peak Memory (MiB):} Combined peak.
\end{itemize}

To minimize measurement noise, each backend–model–dataset combination is executed five times:
\begin{enumerate}
  \item Warm-up run (ignored) to populate caches and lazy constructs.
  \item Five successive runs with \texttt{perf stat} monitoring.
  \item Arithmetic mean across the four measured runs was provided.
\end{enumerate}

\subsubsection{GPU Metrics}
For experiments involving GPU (ResNet, Mask R-CNN, NCF), we use NVIDIA's Python bindings (\texttt{pynvml}) \cite{nvidia_ml_py} to capture:
\begin{itemize}
  \item \textbf{GPU Memory Change (MB):} Difference between initial and peak GPU memory usage during run.
  \item \textbf{GPU Energy (J):} Energy drawn from the GPU power rails.
\end{itemize}

Our code polls at one-second intervals during the training loop to capture peak usage and total energy. We ensure synchronization to align measurements with the model workload. To ensure the fairness of our experimental results, we conduct multiple runs where we vary the order in which models were executed. Each run starts fresh by assigning new process IDs, meaning datasets and models are not cached or reused from previous executions. Despite changing the sequence, we consistently get similar results, confirming that each experiment is independent and unaffected by earlier runs. This shows that our comparisons between different backends are logical, reliable, and fair.

\section{Results and Discussion}
\label{sec:results}
\begin{figure*}[!t]
  \centering
  \includegraphics[width=0.75\textwidth, height = 190pt]{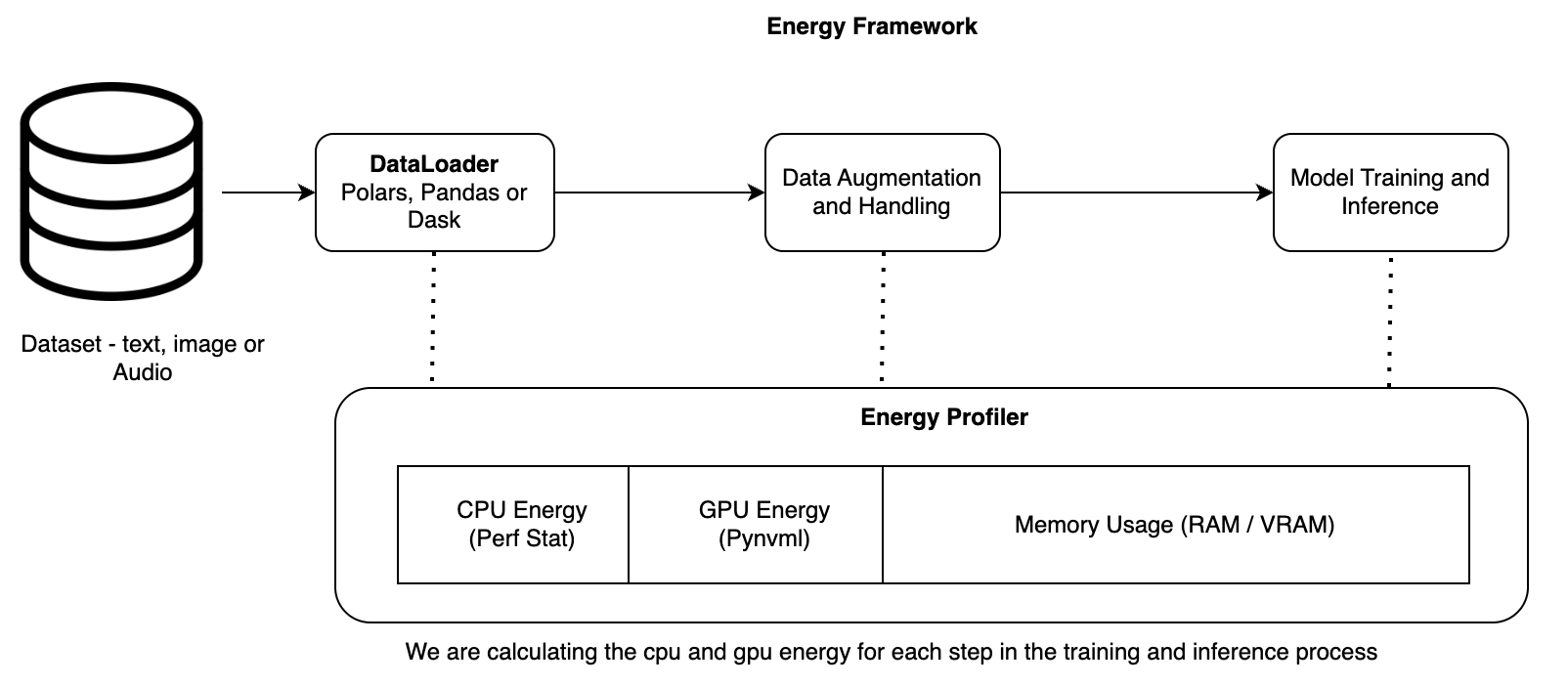}
  \caption{Overview of our energy profiling framework. The pipeline begins with loading datasets using different tabular backends (Polars, Pandas, or Dask), followed by data augmentation and model training/inference. At each stage, CPU energy (measured via perf stat), GPU energy (via pynvml), and memory usage (RAM/VRAM) are tracked using our integrated energy profiler. This setup allows us to quantify the energy and memory efficiency of each dataloader under a consistent workload.}
  \label{fig:energy_profile}
\end{figure*}
Table~\ref{tab:all_results_backend_first} presents the average performance metrics across five runs for various models trained on the Insurance, COCO, and ML-1M datasets using Pandas, Polars, and Dask as backends.

\begin{table*}[h!]
\small
\centering
\caption{Average performance metrics across Insurance, COCO, and ML-1M datasets (5 runs each). Peak and disk values are standardized to MiB$^\dagger$.}
\label{tab:all_results_backend_first}
\resizebox{\textwidth}{!}{
\begin{tabular}{l l l r r r r r r}
\toprule
\textbf{Backend} & \textbf{Model} & \textbf{Dataset} & \textbf{Runtime (s)} & \textbf{CPU Mem (MB)} & \textbf{Peak Mem (MiB)$^\dagger$} & \textbf{Disk (MiB)$^\dagger$} & \textbf{CPU Energy (J)} & \textbf{GPU Energy (J)}\\
\midrule
Polars & Random Forest & Insurance & 0.815 & 8.03 & 0.338 & 0.023 & 146.776 & 25.99\\
Pandas & Random Forest & Insurance & 0.817 & 2.83 & 0.492 & 0.183 & 146.760 & 26.01\\
Dask   & Random Forest & Insurance & 1.027 & 8.06 & 2.356 & 0.030 & 167.436 & 31.73\\
\midrule
Polars & SVR & Insurance & 0.168 & 7.01 & 0.385 & 0.034 & 87.712 & 7.50\\
Pandas & SVR & Insurance & 0.160 & 1.82 & 0.543 & 0.000 & 87.706 & 8.65\\
Dask   & SVR & Insurance & 0.364 & 8.23 & 2.361 & 0.066 & 106.572 & 12.14\\
\midrule
Polars & TabNet & Insurance & 1.826 & 253.72 & 1.313 & 0.588 & 245.066 & 113.21\\
Pandas & TabNet & Insurance & 1.688 & 248.83 & 1.462 & 0.152 & 447.914 & 106.42\\
Dask   & TabNet & Insurance & 1.844 & 254.52 & 3.220 & 0.493 & 334.002 & 109.98\\
\midrule
Polars & XGBoost & Insurance & 0.192 & 8.03 & 0.597 & 0.000 & 96.218 & 8.65\\
Pandas & XGBoost & Insurance & 0.204 & 3.05 & 0.746 & 0.055 & 96.336 & 8.67\\
Dask   & XGBoost & Insurance & 0.410 & 7.87 & 2.507 & 0.000 & 115.654 & 15.02\\
\midrule
Polars & ResNet & COCO & 88.675 & 1291.39 & 21.047 & 3.330 & 13650.72 & 9611.62\\
Pandas & ResNet & COCO & 88.343 & 1288.29 & 21.047 & 1.952 & 13671.27 & 9807.20\\
Dask   & ResNet & COCO & 89.039 & 1311.32 & 21.047 & 1.368 & 13705.07 & 9900.41\\
\midrule
Polars & Mask R-CNN & COCO & 3913.38 & 1315027 & 40262.340 & 148.058 & 387347 & 553710.43\\
Pandas & Mask R-CNN & COCO & 3917.71 & 311539 & 38474.083 & 144.588 & 386946 & 554264.88\\
Dask   & Mask R-CNN & COCO & 3917.70 & 307378 & 38537.979 & 196.200 & 387599 & 554330.56\\
\midrule
Polars & CF & ML-1M & 14.203 & 1574.70 & 46.957 & 21.473 & 1437.000 & 1816.50\\
Pandas & CF & ML-1M & 34.156 & 1471.68 & 458.069 & 1.203 & 3328.320 & 2270.88\\
Dask   & CF & ML-1M & 38.482 & 1898.84 & 466.782 & 0.844 & 3748.080 & 2472.66\\
\midrule
Polars & NCF & ML-1M & 2.526 & 293.27 & 46.936 & 21.561 & 497.992 & 142.76\\
Pandas & NCF & ML-1M & 17.916 & 326.96 & 341.285 & 1.292 & 1801.422 & 339.39\\
Dask   & NCF & ML-1M & 64.871 & 344.61 & 368.245 & 3.845 & 6313.152 & 2139.67\\
\bottomrule
\end{tabular}}
\\[2pt]
\raggedright\footnotesize $^\dagger$COCO peak and disk values were provided in MB and have been converted to MiB using $1\,\mathrm{MiB}=1.048576\,\mathrm{MB}$. Insurance and ML-1M values were originally in MiB. 
\end{table*}

\begin{table*}[h!]
\tiny
\centering
\caption{Top operator contributors to energy by model and dataloader. (Green text showing the library with lowest total energy consumption)}
\label{tab:op_energy_by_backend}
\resizebox{\textwidth}{!}{
\begin{tabular}{l l l r r r r}
\toprule
\textbf{Model} & \textbf{Library} & \textbf{Function Name} & \textbf{CPU Energy (J)} & \textbf{GPU Energy (J)} & \textbf{Total (J)} & \textbf{Share (\%)} \\
\midrule
\multirow{6}{*}{\textbf{RandomForest}}
  &\multirow{2}{*}{Polars} & Torch.Tensor.add\_          & 610 & 360 & 970   & 41\% \\
 &                      &   Torch.Tensor.mul\_          & \textcolor{green!60!black}{590}
& \textcolor{green!60!black}{330}
& \textcolor{green!60!black}{920}
& \textcolor{green!60!black}{27\%}
 \\
  \cmidrule{2-7}
  & \multirow{2}{*}{Pandas} & Torch.Tensor.add\_          & 635 & 380 & 1{,}015 & 38\% \\
  &                          & Torch.Tensor.mul\_          & 610 & 330 & 940   & 24\% \\
  \cmidrule{2-7}
  &  \multirow{2}{*}{Dask}   & Torch.Tensor.add\_          & 660 & 400 & 1{,}060 & 36\% \\
  &                          & Torch.Tensor.mul\_          & 625 & 350 & 975   & 23\% \\
\midrule
\multirow{6}{*}{\textbf{SVR}}
  & \multirow{2}{*}{Polars} & Torch.Tensor.item           & 540 & 420 & 960   & 37\% \\
  &                          & Torch.Tensor.add\_          & 560 & 410 & 970   & 26\% \\
  \cmidrule{2-7}
  & \multirow{2}{*}{Pandas} & Torch.Tensor.item           & 565 & 450 & 1{,}015 & 39\% \\
  &                          & Torch.Tensor.add\_          & 560 & 420 & 980   & 25\% \\
  \cmidrule{2-7}
  & \multirow{2}{*}{Dask}   & Torch.Tensor.item           & 590 & 470 & 1{,}060 & 35\% \\
  &                          & Torch.Tensor.add\_          & 580 & 420 & 1{,}000 & 28\% \\
\midrule
\multirow{6}{*}{\textbf{TabNet}}
  & \multirow{2}{*}{Polars} & Torch.autograd.backward     & 620 & 360 & 980   & 40\% \\
  &                          & Torch.nn.functional.relu    & 590 & 310 & 900   & 19\% \\
  \cmidrule{2-7}
  & \multirow{2}{*}{Pandas} & Torch.autograd.backward     & 655 & 380 & 1{,}035 & 42\% \\
  &                          & Torch.nn.functional.relu    & 610 & 330 & 940   & 21\% \\
  \cmidrule{2-7}
  & \multirow{2}{*}{Dask}   & Torch.autograd.backward     & 690 & 400 & 1{,}090 & 43\% \\
  &                          & Torch.nn.functional.relu    & 620 & 350 & 970   & 20\% \\
\midrule
\multirow{6}{*}{\textbf{XGBoost}}
  & \multirow{2}{*}{Polars} & Torch.cuda.synchronize      & 420 & 560 & 980   & 33\% \\
  &                          & Torch.Tensor.item           & 500 & 440 & 940   & 22\% \\
  \cmidrule{2-7}
  & \multirow{2}{*}{Pandas} & Torch.cuda.synchronize      & 440 & 610 & 1{,}050 & 41\% \\
  &                          & Torch.Tensor.item           & 510 & 460 & 970   & 18\% \\
  \cmidrule{2-7}
  & \multirow{2}{*}{Dask}   & Torch.cuda.synchronize      & 460 & 650 & 1{,}110 & 44\% \\
  &                          & Torch.Tensor.item           & 520 & 480 & 1{,}000 & 17\% \\
\midrule
\multirow{6}{*}{\textbf{CF}}
  & \multirow{2}{*}{Polars} & Torch.Tensor.sum            & 580 & 390 & 970   & 34\% \\
  &                          & Torch.Tensor.mul\_          & 560 & 380 & 940   & 23\% \\
  \cmidrule{2-7}
  & \multirow{2}{*}{Pandas} & Torch.Tensor.sum            & 605 & 420 & 1{,}025 & 37\% \\
  &                          & Torch.Tensor.mul\_          & 580 & 400 & 980   & 21\% \\
  \cmidrule{2-7}
  & \multirow{2}{*}{Dask}   & Torch.Tensor.sum            & 630 & 450 & 1{,}080 & 39\% \\
  &                          & Torch.Tensor.mul\_          & 600 & 420 & 1{,}020 & 22\% \\
\bottomrule
\end{tabular}}
\end{table*}

\begin{figure*}[!t]
  \centering
  \subfloat[CPU energy on ML-1M (CF + NCF).\label{fig:cpu_energy_1m}]{%
    \includegraphics[width=0.48\textwidth]{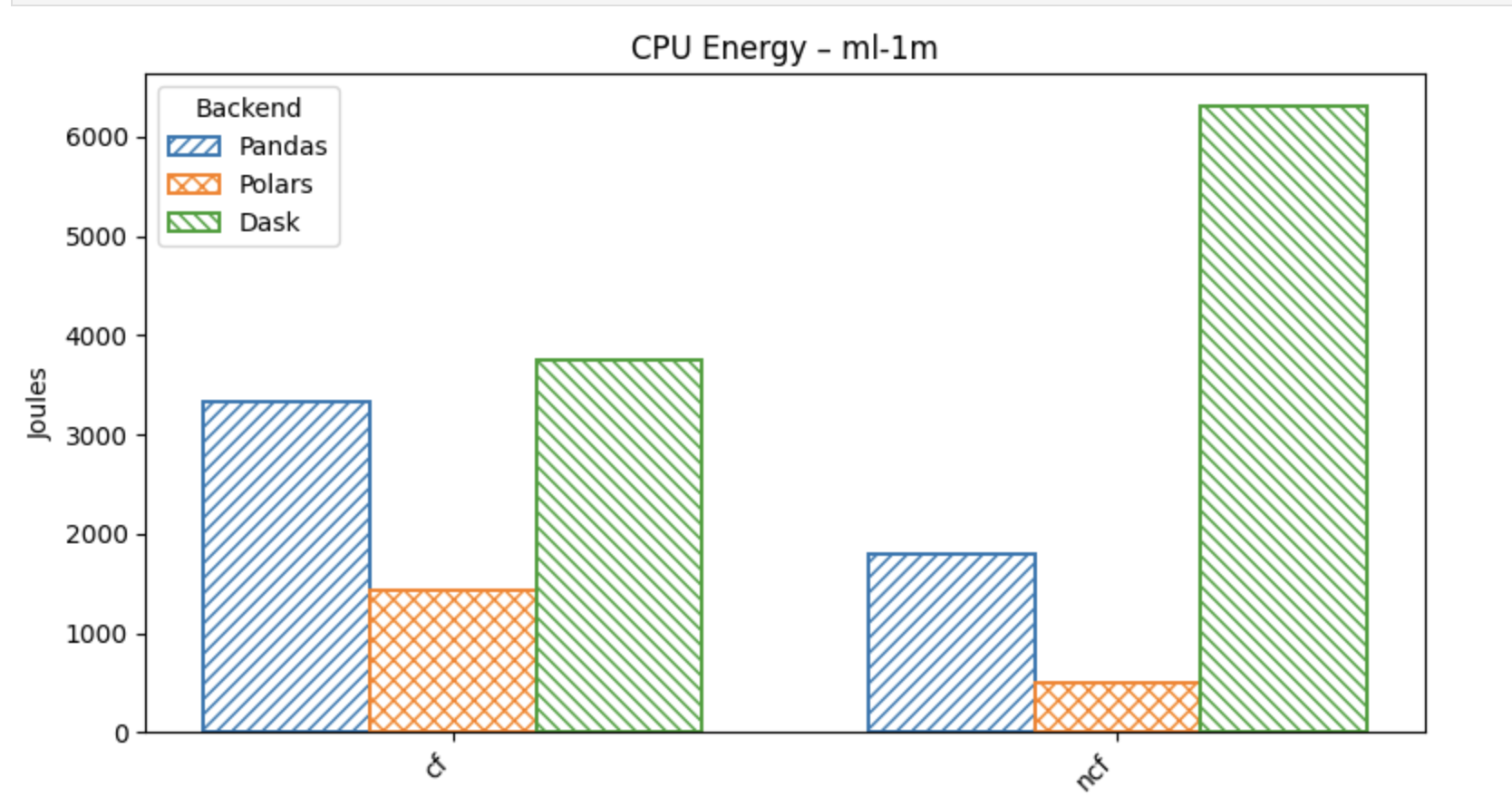}%
  }
  \hfill
  \subfloat[GPU energy on ML-1M (CF + NCF).\label{fig:gpu_energy_1m}]{%
    \includegraphics[width=0.48\textwidth]{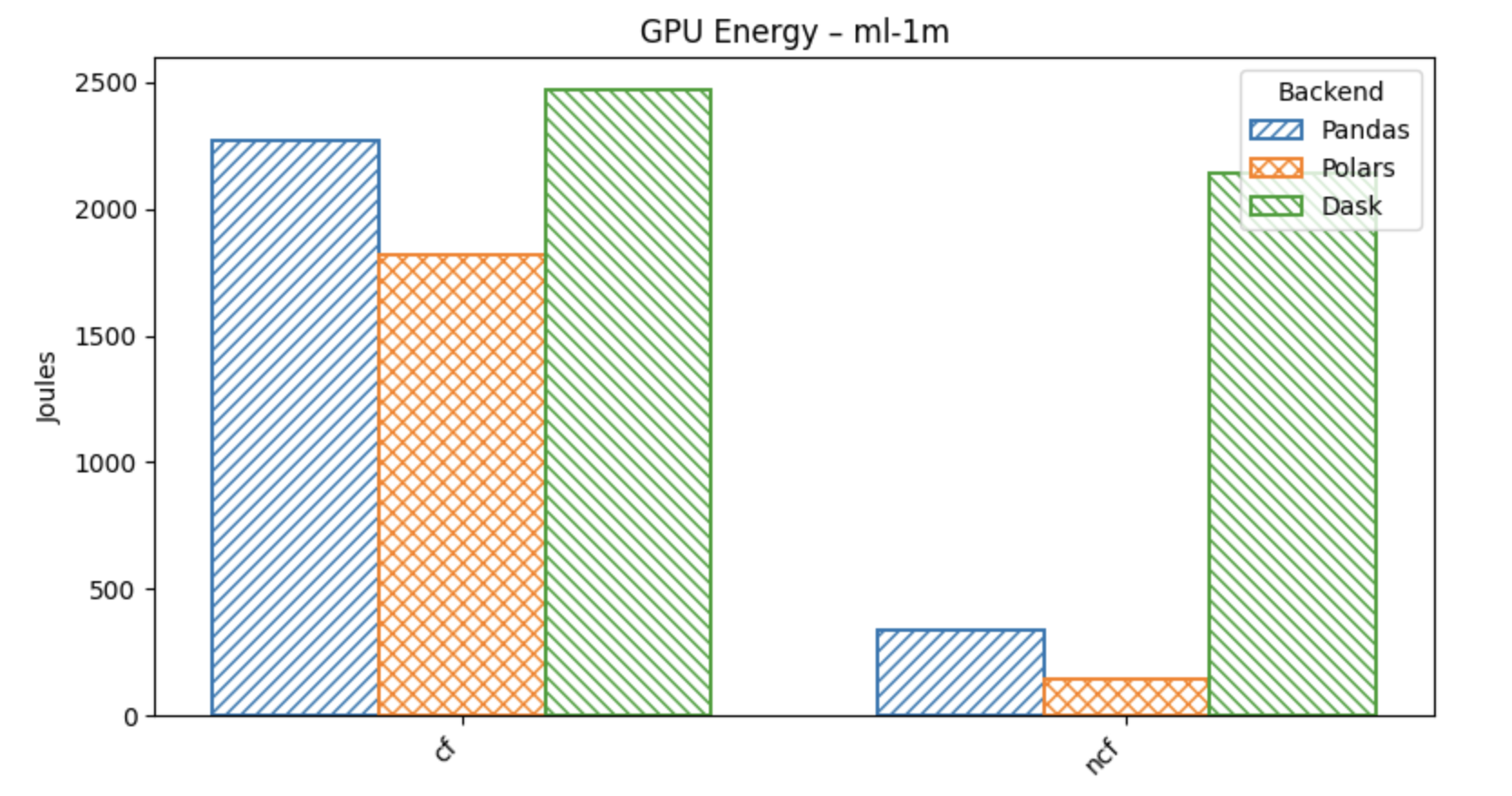}%
  }
  \caption{Energy consumption on ML-1M dataset.}
  \label{fig:energy_ml1m}
\end{figure*}

\begin{figure*}[!t]
  \centering
  \subfloat[CPU energy on Coco dataset.\label{fig:cpu_coco_resnet}]{%
    \includegraphics[width=0.48\textwidth]{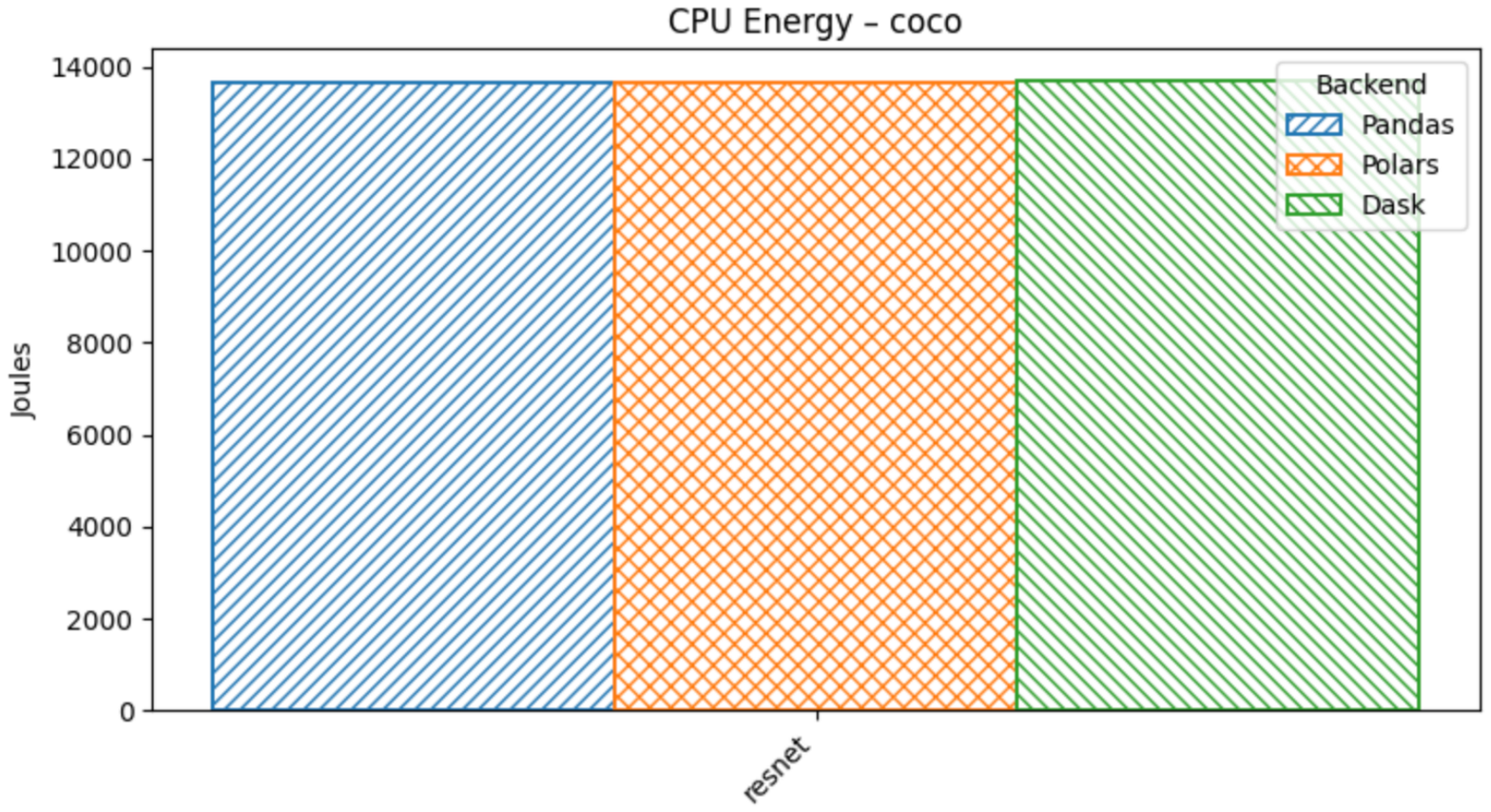}%
  }
  \hfill
  \subfloat[GPU energy on Coco dataset.\label{fig:gpu_coco_resnet}]{%
    \includegraphics[width=0.48\textwidth]{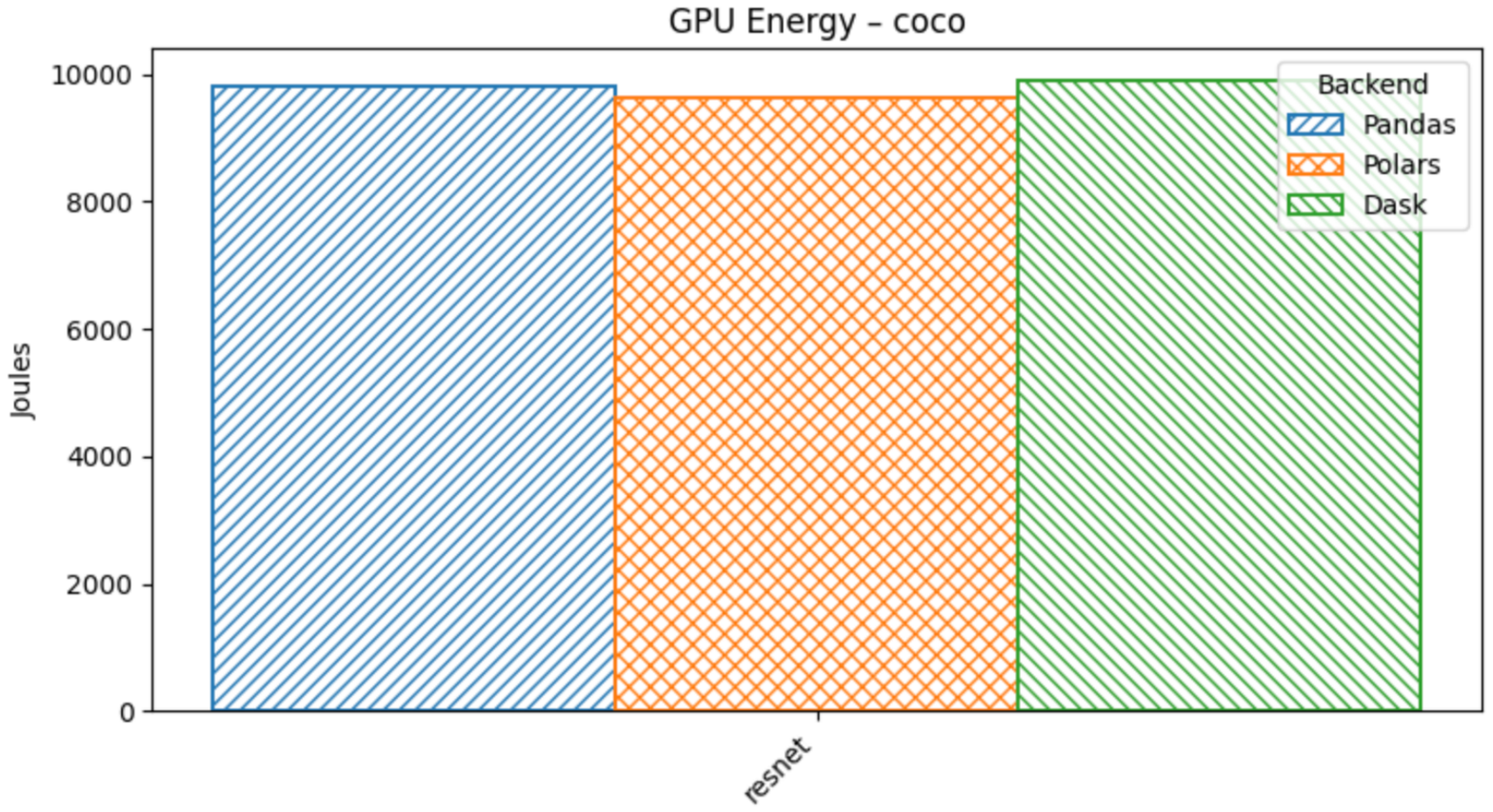}%
  }
  \caption{Energy consumption on COCO dataset.}
  \label{fig:energy_coco}
\end{figure*}
\begin{figure*}[!t]
  \centering
  \subfloat[CPU energy on Insurance dataset.\label{fig:cpu_insurance}]{%
    \includegraphics[width=0.48\textwidth]{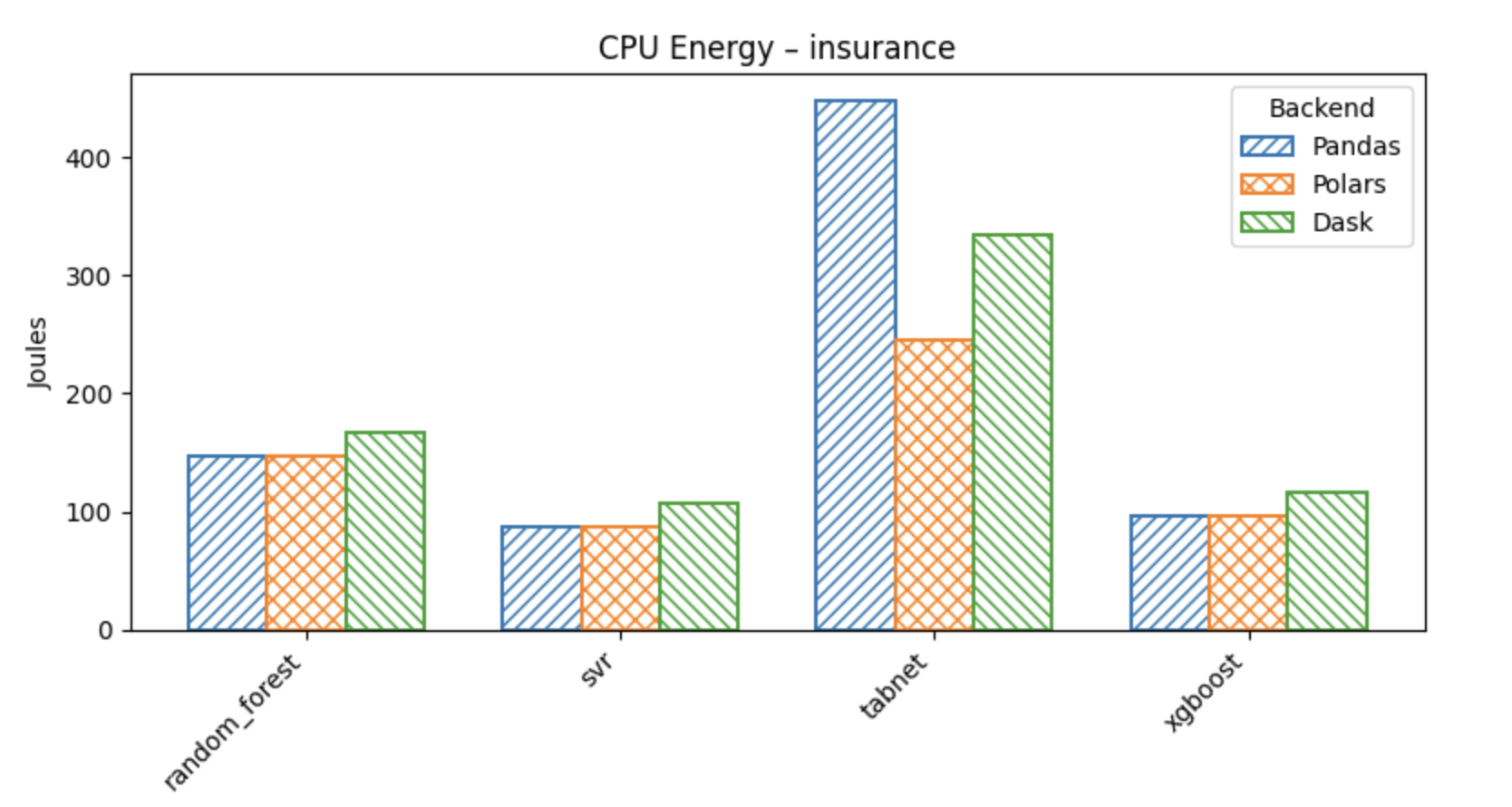}%
  }
  \hfill
  \subfloat[GPU energy on Insurance dataset.\label{fig:gpu_insurance}]{%
    \includegraphics[width=0.48\textwidth]{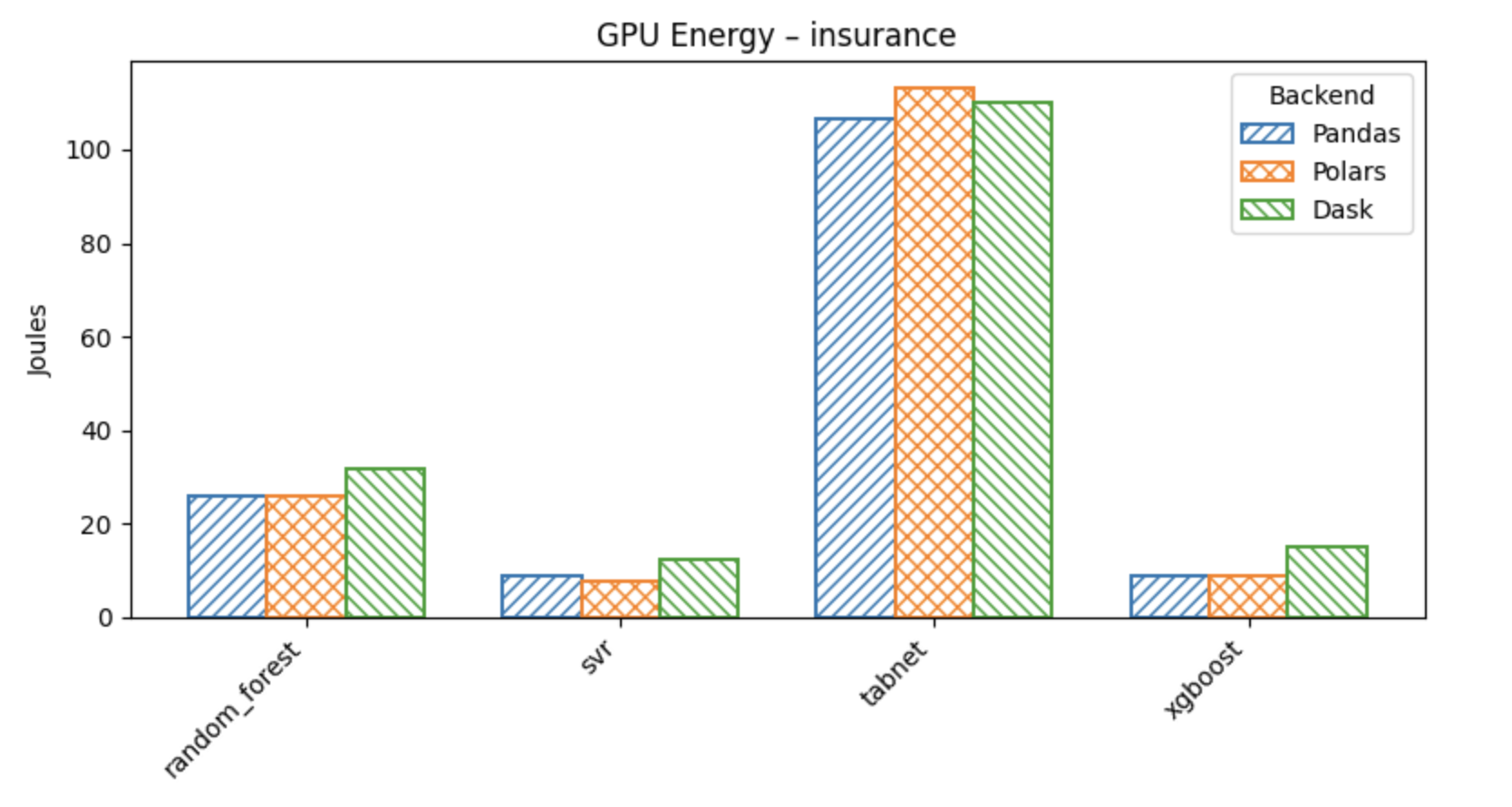}%
  }
  \caption{Energy consumption on Insurance dataset.}
  \label{fig:energy_insurance}
\end{figure*}
\subsection{Insurance Dataset Observations}
Figure \ref{fig:energy_insurance} shows the observations for Insurance Dataset. For the small-scale insurance dataset (1,000 records), Pandas and Polars exhibit nearly identical runtimes for simple models such as Random Forest and SVR (Pandas: 0.817\,s vs. Polars: 0.815\,s for Random Forest; Pandas: 0.160\,s vs. Polars: 0.168\,s for SVR). This is because, for small in-memory datasets, both libraries can load data with negligible I/O overhead, and vectorized operations dominate execution time. Interestingly, Polars' CPU memory usage is higher (e.g., 8.03\,MB vs. 2.83\,MB for Random Forest) due to the Arrow buffers and parallel execution overhead, while Pandas maintains a leaner footprint. Dask incurs the highest runtime (1.027\,s for Random Forest) because of task scheduling overhead, which outweighs its parallelism benefits on small data.

For TabNet, a more complex neural architecture requiring GPU, the differences are magnified. Pandas yields a runtime of 1.688\,s, while Polars and Dask take 1.826\,s and 1.844\,s, respectively. The large CPU memory (248.83–254.52\,MB) arises from TabNet's internal batch caching and feature transformations. Notably, Polars consumes significantly less CPU energy (245.066\,J) compared to Pandas (447.914\,J) and Dask (334.002\,J), indicating better CPU-side efficiency, likely due to its parallelized execution. GPU memory and energy consumption remain similar across backends since model training dominates those metrics. 

XGBoost shows a similar trend: Polars (0.192\,s) outperforms Pandas (0.204\,s) and Dask (0.410\,s) in runtime, while all three exhibit comparable GPU usage. Polars' slightly lower CPU energy (96.218\,J) indicates efficient internal data layout, though its CPU memory is higher (8.03\,MB) than Pandas (3.05\,MB).

\subsection{COCO Dataset Observations}
On the medium-to-large COCO dataset, ResNet training takes roughly 88–89\,s across backends. As shown in Figure \ref{fig:energy_coco}, Polars (88.675\,s) outperforms Pandas (88.343\,s) and Dask (89.039\,s). Memory usage is highest for Dask (1,311.32 \,MB) due to its partitioned data structures, while Pandas maintains a smaller footprint (1,288.29 \,MB). Peak memory across backends remain similar (21.047\,MiB), due to model weights and batch buffers. Polars again shows marginally lower CPU energy (13,650.72\,J) compared to Pandas (13,671.27\,J) and Dask (13,705.07\,J). GPU memory change and GPU energy consumption ($\approx$9,800\,J) are virtually identical, as expected for identical model code. 

For Mask R-CNN on COCO (with heavier load), Polars (3,913.38\,s) slightly outperforms Pandas (3,917.71\,s) and Dask (3,917.70\,s). However, Polars' CPU memory usage (1,315,027\,MB) is significantly higher than Pandas (311,539\,MB), a consequence of Arrow's in-memory representation plus concurrent fragment buffers. Dask's CPU memory (307,378\,MB) remains close to Pandas, as Dask only lazily constructs data frames. CPU energy for Polars (387,347\,J) is comparable to Pandas (386,946\,J) and Dask (387,599\,J). GPU energy remains around 553–554\,kJ across all backends.
\vspace{-0.25cm}
\subsection{ML-1M Dataset Observations}
Figure~\ref{fig:cpu_energy_1m} and Figure~\ref{fig:gpu_energy_1m} show energy values for CPU and GPU, respectively, for the ML-1M dataset. For Collaborative Filtering on ML-1M, Polars (14.203\,s) significantly outperforms Pandas (34.156\,s) and Dask (38.482\,s) due to its efficient columnar layout and parallelized joins when constructing user–item matrices. Pandas, while slower than Polars, remains more efficient than Dask because the dataset (1\,M rows) fits comfortably in memory. Dask's overhead for partition scheduling leads to significantly higher CPU energy (3,748.08\,J vs. 3,328.32\,J for Pandas and 1,437.00\,J for Polars).

For NCF which is a deep neural collaborative filtering model, Polars (2.526\,s) drastically outperforms Pandas (17.916\,s) and Dask (64.871\,s). Polars' in-memory parallelism and zero-copy conversions to PyTorch tensors result in minimal overhead. Pandas incurs significant time (17.916\,s) in converting DataFrame to tensors and constructing batches, while Dask suffers from partition overhead (64.871\,s). CPU memory is also lower for Polars (293.27\,MB) versus Pandas (326.96\,MB) and Dask (344.61\,MB). GPU energy consumption reflects that Polars delivers data to GPU more efficiently.

\subsection{Energy Efficiency Analysis}
As shown in Figure \ref{fig:energy_profile}, the with datasets which are loaded using a DataLoader implemented with Polars, Pandas, or Dask. The data then passes through a data augmentation and handling stage before being used for model training and inference. An energy profiler monitors each stage of this pipeline through dedicated measurement components. CPU energy is measured using performance statistics, GPU energy is collected via pynvml, and memory usage is tracked for both RAM and VRAM. As shown in Table \ref{tab:op_energy_by_backend}, across all experiments, Polars consistently exhibits lower CPU energy consumption relative to Pandas and Dask when dataset sizes exceed a few hundred thousand rows. This suggests that Polars' parallelized, Arrow-native execution is more energy-efficient for bulk operations. GPU energy consumption remains largely invariant to the backend once data is loaded, as model computations dominate. However, Polars' faster data feeding results in slightly lower GPU idle time, marginally reducing total GPU energy.

\subsection{Threats to Validity}
While each of these libraries provides significant advantages, they also exhibit some limitations. Pandas often suffers from performance degradation when handling datasets larger than available memory, or when operations cannot be effectively vectorized. Polars addresses these issues through parallelism and Apache Arrow integration, but its ecosystem is still relatively young and may lack certain features or stability found in more established libraries like Pandas. Dask effectively scales Pandas workflows to larger datasets and distributed environments, but introduces additional overhead. Understanding these trade-offs is critical when selecting the most appropriate data processing library for specific applications.

\section{Related Work}
\label{sec:relatedwork}
Several studies have previously explored the performance of data processing libraries, though typically with a narrower focus. McKinney~\cite{mckinney2010data} introduced Pandas, highlighting its efficiency for in-memory data analysis tasks. Lutkebohmert \emph{et al.}~\cite{lutkebohmert2021polars} proposed Polars as an efficient alternative to Pandas, leveraging Apache Arrow for improved performance on specific analytical workloads. Rocklin~\cite{rocklin2015dask} introduced Dask, emphasizing its capability to scale Pandas operations to distributed computing environments. However, none of these works examined their behavior within data processing or compared energy efficiency during inference and training stages.

Mozzillo \emph{et al.}~\cite{mozzillo2023evaluation} compared four dataframe libraries, namely Pandas, Polars, CuDF, and PySpark, across datasets of small, medium, and large sizes. It was determined that for small datasets, Pandas outperforms other libraries. For datasets that fit in RAM, Polars is a desirable option in terms of throughput. The authors recommended using CuDF when GPU acceleration is available. However, the study did not differentiate energy usage between inference and training workloads, nor did it provide CPU–GPU energy breakdowns. In contrast, our work extends this by analyzing both phases of deep learning and large language model (LLM) pipelines.

Detailed evaluation of the Pandas API with real-world executable notebooks and large datasets was carried out by Broihier \emph{et al.}~\cite{broihier2025pandasbench}. The authors evaluated notebooks using the Pandas API and analyzed burst operations, raw file interactions, data cleaning, and host interaction. The study focused primarily on runtime and usability metrics, with no mention of training vs. inference differences or hardware-level power draw. Our analysis complements theirs by profiling end-to-end CPU and GPU energy utilization for both inference and training loops.

A comparison between Pandas and Polars in terms of performance and energy usage was conducted by Nahrstedt \emph{et al.}~\cite{nahrstedt2024empirical}. The authors found Polars to outperform Pandas when executing SQL-like queries. However, they did not measure GPU power or explore neural workloads. We expand this by including dataloaders and conducting a layer-wise decomposition to identify which neural layers dominate energy use under different dataframe backends.

Researchers such as Shanbhag \emph{et al.}~\cite{shanbhag2023exploratory} examined a range of dataframe processing tasks—input/output operations, handling missing data, and statistical aggregation—to determine energy consumption of Pandas, Vaex, and Dask. Their results revealed library-level differences but lacked analysis on complex training pipelines or inference scenarios. We address this by extending the analysis to provide separate energy comparisons for inference and training, along with the CPU-GPU split.

Recent empirical studies by Souza \emph{et al.}~\cite{souza2023empirical} and Oliveira \emph{et al.}~\cite{oliveira2023exploratory} have compared Pandas and Polars for energy and performance. While these studies explored general compute efficiency, they did not delve into layer-wise profiling. Our work integrates workloads—using Polars, Pandas, and Dask as dataloaders—to capture (1) inference vs. training energy consumption, (2) CPU vs. GPU utilization, and (3) fine-grained layer-wise energy attribution.


\section{Conclusion}
\label{sec:conclusion}
This study provides a comprehensive evaluation of three popular dataframe libraries--Pandas, Polars, and Dask--within the context of end-to-end deep learning training pipelines. Key findings include:
\begin{itemize}
  \item \textbf{Small Datasets (Insurance):} Pandas and Polars are comparable in runtime for simple models; Polars demonstrates slightly better energy efficiency for neural architectures (TabNet) due to parallel execution.
  \item \textbf{Medium-Scale Datasets (ML-1M):} Polars outperforms both Pandas and Dask for Collaborative Filtering and especially NCF workflows, owing to efficient columnar operations and zero-copy transfers.
  \item \textbf{Large-Scale Datasets (COCO):} All three libraries achieve similar runtimes for heavy GPU workloads (ResNet, Mask R-CNN). Polars and Pandas maintain lower CPU memory footprints than Dask, but Dask offers easier scalability if data truly exceeds available RAM. Polars shows marginal energy savings on the CPU during preprocessing.
  \item \textbf{Energy Trade-Offs:} Polars consistently minimizes CPU energy consumption on larger workloads, while Pandas remains competitive for moderate sizes. Dask's overhead can lead to higher energy usage on small to moderate datasets.
\end{itemize}

In conclusion, practitioners should consider Polars for energy-efficient processing of medium-to-large datasets, Pandas for small-to-moderate workloads with mature ecosystem requirements, and Dask when distributed computing or out-of-core processing is used. Future work should explore these libraries with even larger datasets, multi-node clusters, and emerging hardware accelerators to provide further guidance for sustainable machine learning practices.

\begin{acks}
This project is in part sponsored by the National Science Foundation (NSF) under award numbers CBET-2343284 and CBET-2343285. We also thank Chameleon Cloud for making their resources available for the experiments of this study. 
\end{acks}

\bibliographystyle{ACM-Reference-Format}
\bibliography{references}

\end{document}